# High quality $Fe_{1+y}Te$ synthesized by chemical vapor deposition with conspicuous vortex flow

*Lu Lv, Lihong Hu, Weikang Dong, Jingyi Duan, Ping Wang, Peiling Li, Fanming Qu, Li Lu, Zimeng Ye, Junhao Zhao, Jiafang Li, Fang Deng, Guangtong Liu\*, Jiadong Zhou\*, Yanfeng Gao\**

L. Lv, Prof. Y. Gao.

School of Materials Science and Engineering, Shanghai University, Shanghai 200444, China

Email: yfgao@shu.edu.cn

L. Lv, W. Dong, J. Duan, P. Wang, Ms Z. Ye, J. Zhao, Prof. J. Li, Prof. F. Deng, Prof. J. Zhou

Centre for Quantum Physics, Key Laboratory of Advanced Optoelectronic Quantum Architecture and Measurement (MOE), School of Physics, Beijing Institute of Technology, Beijing 10081, China

Email: jdzhou@bit.edu.cn

L. Hu, Prof. P. Li, Prof. F. Qu, Prof. L. Lu, Prof. G. Liu

Beijing National Laboratory for Condensed Matter Physics, Institute of Physics, Chinese Academy of Sciences, Beijing 100190, China

Email: gtliu@iphy.ac.cn

Prof. P. Li, Prof. F. Qu, Prof. L. Lu, Prof. G. Liu

Hefei National Laboratory, Hefei 230088, China

Prof. F. Qu, Prof. L. Lu, Prof. G. Liu

Songshan Lake Materials Laboratory, Dongguan, Guangdong 523808, China

Prof. J. Zhou

Advanced Research Institute of Multidisciplinary Science, Beijing Institute of Technology, Beijing 10081, China

Prof. Y. Gao

School of Chemical and Environmental Engineering, Anhui Polytechnic University, Wuhu, Anhui 241000, China

L. Lv, L. Hu and W. Dong contributed equally to this work.

## Abstract

Two-dimensional (2D) materials provide an ideal platform to explore novel superconducting behavior including Ising superconductivity, topological superconductivity and Majorana bound states in different 2D stoichiometric Ta-, Nb-, and Fe-based crystals. However, tuning the element content in 2D compounds for regulating their superconductivity has not been realized. In this work, we report the synthesis of high quality $Fe_{1+y}Te$ with tunable Fe content by chemical vapor deposition (CVD). The quality and composition of $Fe_{1+y}Te$ are characterized by Raman spectroscopy, X-ray photoelectron spectroscopy (XPS) and scanning transmission electron microscopy (STEM). The superconducting behavior of $Fe_{1+y}Te$ crystals with varying Fe contents is observed. The superconducting transition of selected $Fe_{1.13\pm0.06}Te$ sample is sharp ($\Delta T_c = 1$ K), while $Fe_{1.43\pm0.07}Te$ with a high-Fe content shows a relative broad superconducting transition ($\Delta T_c = 2.6$ K) at zero magnetic field. Significantly, the conspicuous vortex flow and a transition from a 3D vortex liquid state to a 2D vortex liquid state is observed in $Fe_{1.43\pm0.07}Te$ sample. Our work highlights the tunability of the superconducting properties of $Fe_{1+y}Te$ and sheds light on the vortex dynamics in Fe-based superconductors, which facilitates us to understand the intrinsic mechanisms of high-temperature superconductivity.

## Introduction

Two-dimensional (2D) materials provide an ideal platform to explore superconductivity in the 2D limit. In recent years, 2D materials such as twisted bilayer graphene,[1-4] monolayer NbSe$_2$,[5-8] TaS$_2$,[9-11] and monolayer FeSe[12-14] have been fully studied, showcasing fascinating physical phenomena such as Ising superconductivity, topological superconductivity and Majorana bound states. Among them, 2D Fe-based superconductors have garnered extensive attention due to their exceptionally high transition temperature ($T_c$) and complex superconducting mechanisms.[15-17] For instance, researches have shown that suppressing antiferromagnetic order by doping enables superconductivity in FeTe$_x$Se$_{1-x}$[18-19] and FeTe$_x$S$_{1-x}$[20-21]. The interface engineering can induce superconductivity in FeTe/Bi$_2$Te$_3$[22-23] and FeTe/SrTiO$_3$[24]. Furthermore, Fe$_{1+x}$Te/Bi$_2$Te$_3$ bilayers[25] exhibit superconductivity attributed to the Fe$_{1+x}$Te layer. We can conclude from above that the previous studies in Fe-based superconductivity mainly focus on the stoichiometric crystals. Tuning the Fe content in Fe-based 2D materials to regulate the superconductivity has not been realized, although the Fe-based 2D materials with a little range Fe content change has been achieved. Therefore, it is urgent to prepare non-stoichiometric Fe-based crystals directly for studying the superconducting behaviors.

Chemical vapor deposition (CVD) has been widely utilized to produce different types of 2D Fe$_a$X$_b$ (X=S,Se,Te), primarily owing to its low-cost and practical industrial scalability.[26] However, most reported studies on 2D Fe$_a$X$_b$ mainly focus on phase modulation or magnetism research[27-40] and no reports on tuning the Fe content to regulate their superconductivity due to the following reasons. Firstly, introducing excessive Fe into layered structures is challenging as it may lead to intercalation structures, which will disrupt inherent physical properties. Secondly, Fe$_a$X$_b$ consists of multiple components, including FeX, FeX$_2$, Fe$_2$X$_3$, Fe$_3$X$_4$, Fe$_7$X$_8$, and so on, making it difficult to control the preparation of single composition crystal. Therefore, synthesizing Fe$_a$X$_b$ materials with tunable Fe content via CVD is the key to studying the impact of Fe content on the physical properties.

Herein, we report the high-quality tetragonal superconducting $Fe_{1+y}Te$ (y = 0~0.43) nanoflakes grown by CVD method under atmospheric pressure with the growth temperature of 520-600 °C. By tuning the growth temperature and the amounts of precursors, the $Fe_{1+y}Te$ with different thickness and y value can be obtained. Furthermore, Raman spectroscopy and X-ray photoelectron spectroscopy (XPS) were employed to verify the quality of nanoflakes. The atomic structure of $Fe_{1+y}Te$ was revealed by high-angle annular dark-field scanning transmission electron microscopy (HAADF-STEM). Interestingly, different superconducting behaviors in $Fe_{1+y}Te$ samples were observed. Specifically, the $Fe_{1.13\pm0.06}Te$ sample exhibits a sharp superconducting transition with the $T_c$ of 10.2 K at a high magnetic field (12 T), while $Fe_{1.43\pm0.07}Te$ shows a relatively broad superconducting transition with a suppressed $T_c$ (12 T) of 3.8 K. Impressively, we observed a transition from a 3D vortex liquid state to a 2D vortex liquid state in $Fe_{1.43\pm0.07}Te$ sample. These findings will help our understanding of the mechanisms underlying high-temperature superconductivity. Overall, our work makes a significant contribution to the field of 2D superconducting materials.

## 2. Results and discussion

### 2.1. Growth and characterization of tetragonal FeTe nanoflakes.

**Figure 1**a provides a detailed schematic of the CVD growth setup with a 1-inch quartz tube. To prepare the FeTe, a porcelain boat loaded with tellurium (Te) powder was located at upstream of the quartz tube, another boat containing $FeCl_2$ precursor was placed in the center of the tube, and a $Si/SiO_2$ wafer was used as substrate. Since $FeCl_2$ powder has a low melting point and tends to volatilize at low temperature, we used large-size $FeCl_2$ particles to reduce the vapor pressure and ultimately facilitate the 2D tetragonal FeTe growth. The $Ar/H_2$ (100/5) gas was used as the carrier gas. The top and side views of the FeTe crystal structure are shown in Figure 1b. It can be seen that the atoms of FeTe are arranged in a tetragonal pattern, belonging to P4/*nmm* space group

with the lattice parameters of a = 3.66 Å, b = 3.66 Å, and c = 6.51 Å, respectively. Figure 1c shows the optical image (OM) of the synthesized FeTe with a size of 5 μm. Atomic force microscopy (AFM) was further used to check the morphology and thickness of grown FeTe. As shown in the AFM image (Figure 1d), ultrathin tetragonal FeTe flake with a thickness of 4.3 nm was observed.

Raman spectroscopy was used to investigate the crystalline quality of FeTe. Figure S1 (Supporting Information) displays the Raman spectrum of tetragonal FeTe with a thickness of 15.5 nm, in which three Raman peaks were observed at 91 cm$^{-1}$, 121 cm$^{-1}$ and 138 cm$^{-1}$. The Raman peak at 138 cm$^{-1}$ originates from the $A_{1g}$ mode, while the peak at 121 cm$^{-1}$ corresponds to the $E_g$ mode. The corresponding Raman mapping at 121 cm$^{-1}$ is shown in Figure 1e and presents a consistent contrast over the surface. A slight enhancement of the contrast along the edges is also observed due to the edge thickness increasing. X-ray photoelectron spectroscopy (XPS) was used to confirm the elemental composition and valence state of the grown samples. The XPS results for the Fe and Te elements in the tetragonal FeTe crystals are shown in Figure 1f and Figure 1g, respectively, and the corresponding full XPS spectra is shown in Figure S2 (Supporting Information). Obviously, the peaks observed at binding energies of about 706.3 eV and 719.2 eV are attributed to $2p_{3/2}$ and $2p_{1/2}$ states of $Fe^{2+}$ (Figure 1f). Besides, the blue peaks observed at 706.6 eV and 719.4 eV in Figure 1f represent $Fe^{3+}$ $2p_{3/2}$ and $Fe^{3+}$ $2p_{1/2}$, respectively, which may originate from slight oxidation of the sample. The peaks observed at binding energies of about 572.4 eV and 582.9 eV are attributed to $3d_{5/2}$ and $3d_{3/2}$ states of $Te^{2-}$, as shown in Figure 1g. All these findings are in good consistent with previously studies.[37, 39]

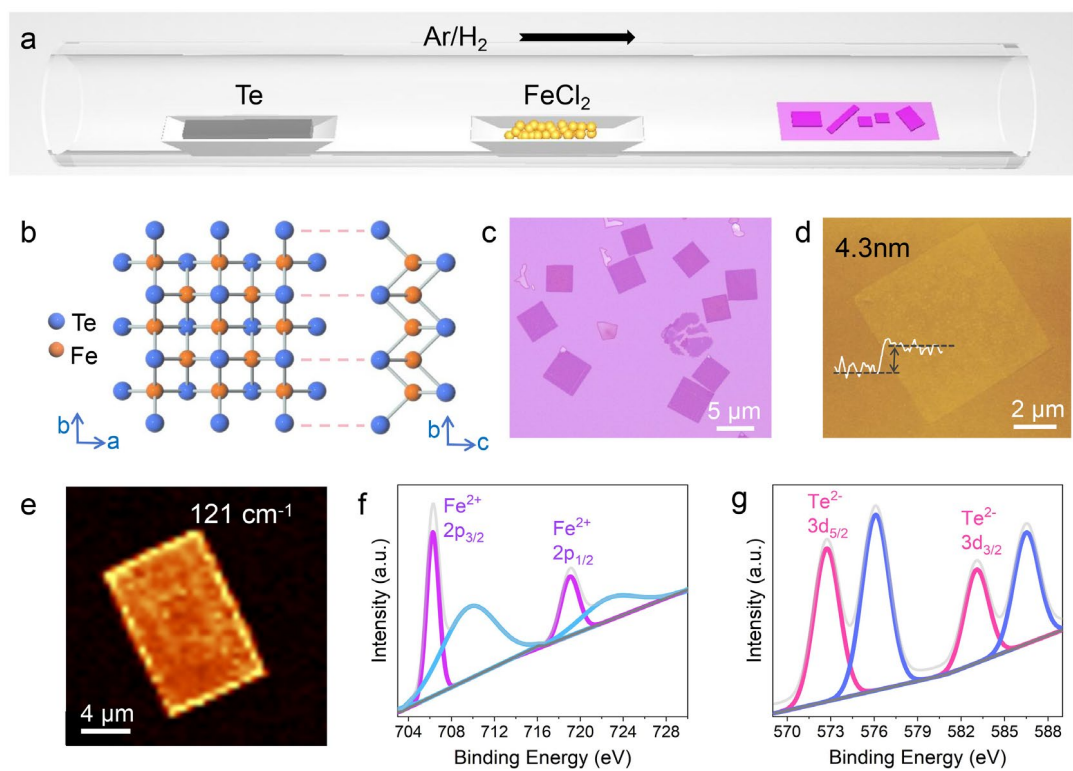

**Figure 1.** a) The schematic of the CVD setup. b) Top view and side view of FeTe crystal structure. Fe atoms are represented by orange balls, and tellurium atoms are represented by blue balls. The lattice constant along a or b axis is 0.366 nm. c) Optical image of as-synthesized FeTe on SiO$_2$/Si substrate. The scale bar is 5 μm. d) The AFM image of FeTe nanoflake, corresponding height profile at the edge shows the thickness of FeTe nanoflake is about 4.3 nm. e) The Raman mapping of a FeTe crystal at 121 cm$^{-1}$. f, g) XPS spectra of Fe and Te elements, respectively.

We further conducted a detailed investigation into the atomic structure of FeTe using High-angle annular dark-field STEM (HAADF-STEM) and energy-dispersive X-ray spectroscopy (EDS) mapping techniques. **Figure 2**a shows a low-magnified cross-section STEM image of the sample obtained after performing a longitudinal cut of the sample using a FEI Helios G4 UC dual-beam microscope. The enlarged STEM image of area 1 is shown in Figure 2b, and the magnified image of the selected region in Figure 2b is shown in Figure 2c. The Fe atoms (gray dots) and Te atoms (white dots) show a sharp contrast due to the different atomic numbers and can be distinguished clearly. We measured the distance between two Fe atoms along the a-axis to be 0.646 nm (Figure 2c), which agrees well with the typical in-plane lattice constant (~0.651 nm) of a FeTe nanosheet (Figure S3b (Supporting Information)). We can also see that there is a good match between the atomic structure and the atomic model shown in Figure 2c. By measuring the distance of 0.39 nm between the two Te atoms along the a-axis (Figure S3c), we further confirmed the agreement of the measured plane spacing with the theoretical value. The clearly selected area electron diffraction (SAED) patterns in Figure 2d confirm the single crystalline nature of the sample. In addition, we performed energy-dispersive X-ray (EDX) elemental mapping (Figure 2e-f) to determine the chemical composition and element distribution of $Fe_{1+y}Te$, which shows an atomic ratio of 50.9: 49.1 for Fe and Te elements (Figure S3a), indicating the stoichiometric ratio of FeTe. Furthermore, we show the temperature dependence of sheet resistance $R_s(T)$ for FeTe (#5) in Figure S4 (Supporting Information).

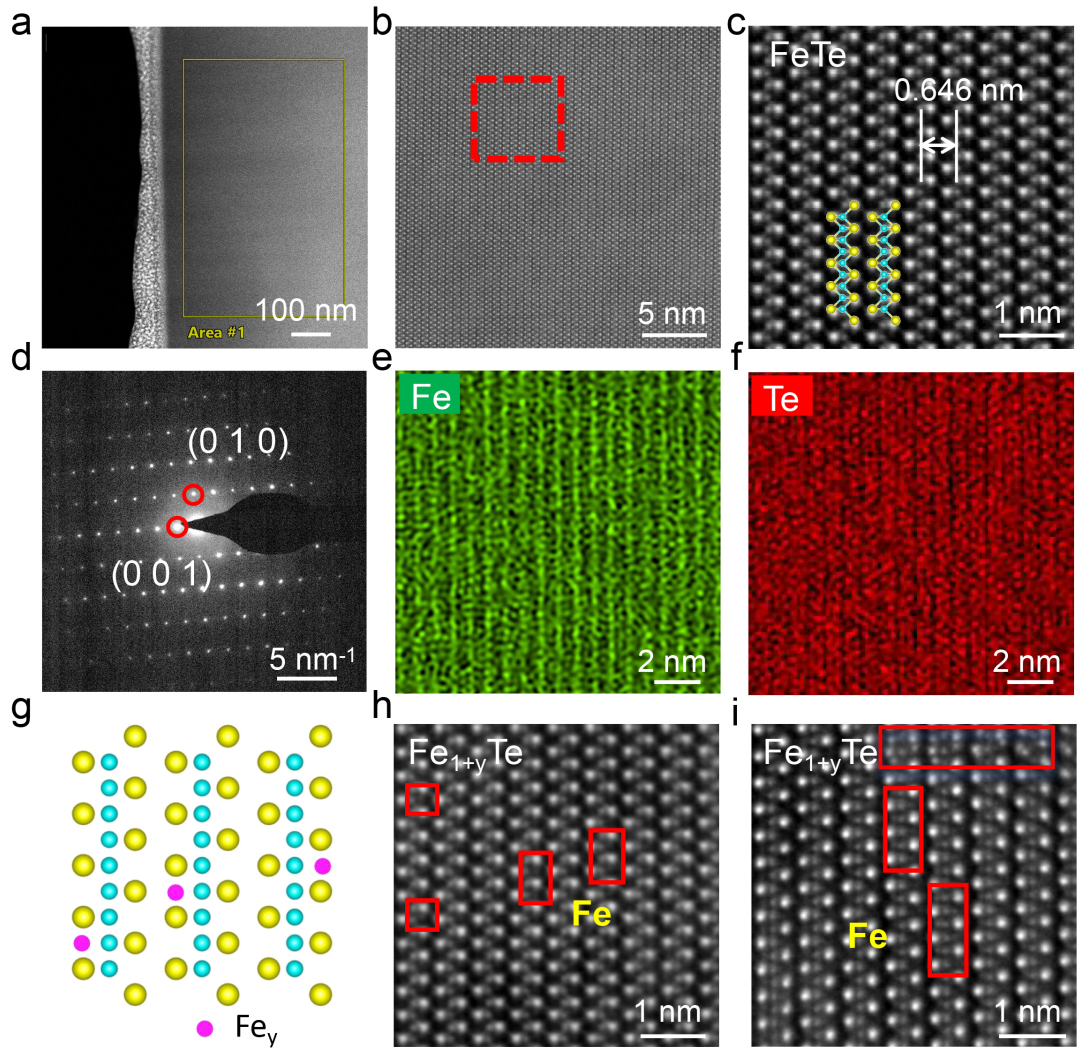

**Figure 2.** a) Low-magnified STEM image of the cross-section of a FeTe nanoflake. b) The HRTEM image of the selected area in Figure 2a. Yellow and blue colors represent Te and Fe atoms, respectively. c) Enlarged STEM image of selected area in Figure 2b. d) Corresponding SAED image of FeTe crystal. e, f) EDX elemental mapping images of Fe and Te, respectively. g) Side view of $Fe_{1+y}Te$ crystal structure. The excess $Fe_y$ atoms are represented by purple balls. h, i) The high-magnified images of $Fe_{1.13\pm0.06}Te$ and $Fe_{1.43\pm0.07}Te$ crystals in [1 0 0] direction, respectively. The red rectangular boxes show the typical areas that have interstitial Fe atoms.

**2.2. Synthesis and modulation of tetragonal $Fe_{1+y}Te$ nanoflakes.**

In addition to synthesize the $Fe_{1+y}Te$ with different Fe content, we also tried to modulate the Fe content to investigate the superconducting properties. It is worth noting that tetragonal FeTe is more stable in Te deficient environment,[37] similar to tetragonal FeSe.[30] Therefore, in order to synthesize tetragonal $Fe_{1+y}Te$ samples with high Fe content, we reduced the amount of Te to create a Fe-rich environment. The Fe/Te atomic ratio is determined by the mass ratio between $FeCl_2$ and Te precursors involved in the reaction, which is influenced by the growth temperature and the distance of two precursors. By adjusting the Fe/Te mass ratio (2:1 ~ 13:1), we can obtain $Fe_{1+y}Te$ nanosheets with different Fe contents. Figure 2g shows the side view of $Fe_{1+y}Te$ crystal structure, where the excess $Fe_y$ atoms represented by purple color balls are distributed randomly in the gap between the two Te atoms. We also conducted STEM characterization on $Fe_{1+y}Te$ with high Fe contents, as shown in Figure 2h and Figure 2i respectively. The corresponding EDX elemental analysis are shown in Figure S5 (Supporting Information) and Figure S6 (Supporting Information). In Figure 2h and Figure 2i, we observed the presence of excess Fe atoms between Te-Te pair in the red rectangular boxes. These are present as interstitial Fe atoms, which is in agreement with previous reports.[41]. To precisely confirm this structure, we conducted STEM characterization along [1 1 0] crystal direction, as shown in Figure S7 (Supporting Information). Raman spectra of $Fe_{1+y}Te$ nanoflakes with different Fe contents are shown in Figure 3l. Here, in order to eliminate the influence of other factors, we only adjusted the distance between two precursors to regulate the Fe content with growth temperature and gas flow constant. Then, we selected samples with almost same thickness for Raman characterization. Notably, with increasing y value in tetragonal $Fe_{1+y}Te$ nanosheets, the blue-shift of $E_g$ Raman peaks is obviously observed, which is possibly caused by the decrease in interlayer spacing with increasing amounts of Fe. There is almost no change in the $A_{1g}$ Raman peak, indicating that Fe elements are well distributed within the layer. We further systematically investigated the effect of synthetic parameters on the growth of $Fe_{1+y}Te$ samples to modulate the thickness, size and the Fe contents of $Fe_{1+y}Te$ nanoflakes. We found that the growth temperature and gas flow rate significantly affect the thickness and morphology of $Fe_{1+y}Te$ nanosheets. We first studied the effect of growth temperature on the thickness of $Fe_{1+y}Te$ nanoflakes with the growth parameters (gas flow, Fe/Te mass ratio and growth time) remaining constant. **Figure 3**a-g display the typical optical images of $Fe_{1+y}Te$ crystals prepared at different temperatures (see Figure S8 (Supporting Information) for details). The corresponding AFM images are shown in Figure S9 (Supporting Information). Briefly, the size and thicknesses of $Fe_{1+y}Te$ can be tuned from ~ 2 to ~ 40 μm and from ~ 4.3 to ~ 53.3 nm, respectively, with the growth temperature increasing from 520 °C to 600 °C. Raman measurements were also performed on $Fe_{1+y}Te$ samples with different thicknesses. As shown in Figure 3b, we note that the thin samples (< 6 nm) show only one Raman peak (158 cm$^{-1}$) and the peak position is slightly red-shifted with thickness increasing, while in the thick samples, the Raman peaks at 121 cm$^{-1}$ and 138 cm$^{-1}$ are constantly shifted to higher frequencies as the thickness increases, which is often observed in the 2D materials.[38] In addition, we explored the effect of flow rate on the growth of $Fe_{1+y}Te$. As shown in Figure 3h-k, the length increases from 3 μm to 250 μm when the flow rate is increased from 80 sccm to 120 sccm. The corresponding Raman spectra of the $Fe_{1+y}Te$ with different edge lengths are shown in Figure 3n.

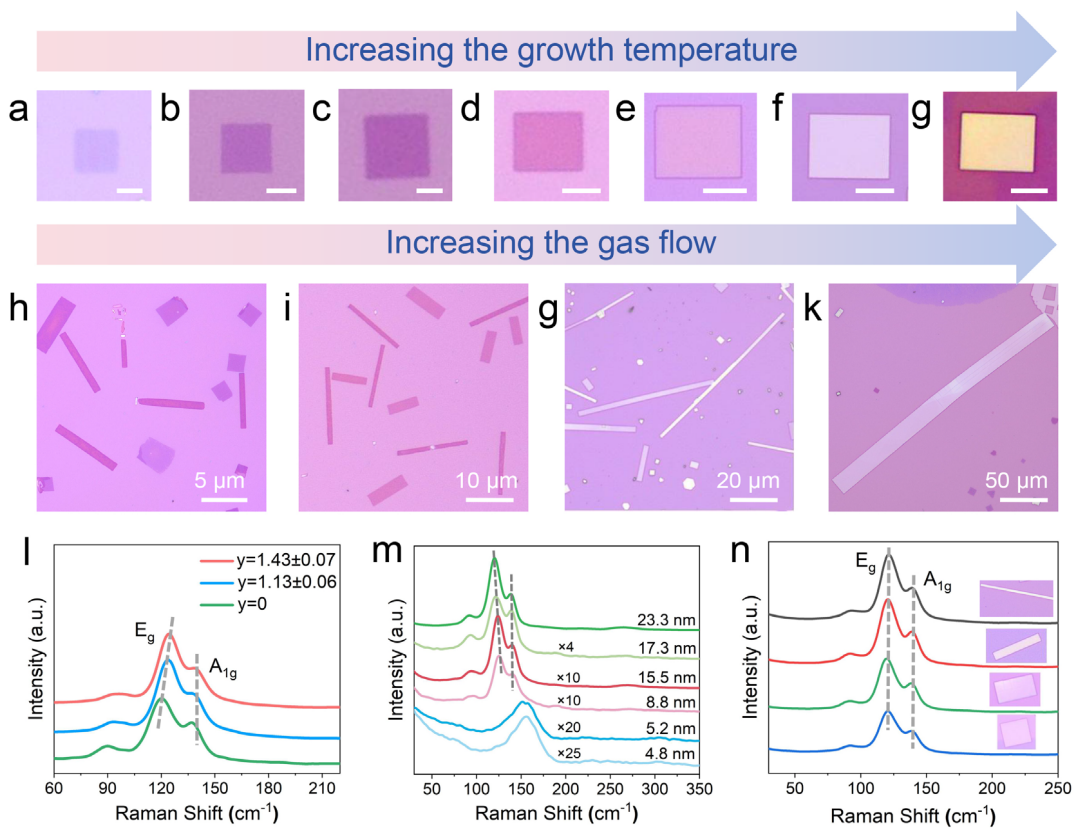

**Figure 3.** a-g) Growth temperature-dependent optical images of $Fe_{1+y}Te$. The scale bar is 2 μm (a), 5 μm (b), 5 μm (c), 10 μm (d), 15 μm (e), 20 μm (f) and 25 μm (g), respectively. h-k) Gas flow-dependent optical images of as-grown $Fe_{1+y}Te$ nanoflakes on $SiO_2$/Si substrates. l) Raman spectra of $Fe_{1+y}Te$ samples with different Fe contents. m) Thickness-dependent Raman spectra of the $Fe_{1+y}Te$ nanoflakes. n) Raman spectra of $Fe_{1+y}Te$ nanoflakes with different edge lengths.

### 2.3. Superconducting characteristics of $Fe_{1.13\pm0.06}Te$ and $Fe_{1.43\pm0.07}Te$.

**Figure 4**a shows the typical temperature dependence of sheet resistance ($R_s$) of $Fe_{1.13\pm0.06}Te$. At high temperatures, the resistance slowly increases with decreasing temperature, exhibiting a semiconductor-like behavior. However, in the vicinity of 80 K, $R_s$ begins to decrease and forms a broad peak. The change in sheet resistance suggests a tetragonal-monoclinic structural transition, accompanied by the formation of an antiferromagnetic order as reported by previous studies.[25, 42-43] As the temperature decreases further, a sharp drop in $R_s$ towards zero is observed at around 12.6 K, indicating the onset of a superconducting transition ($T_{c,\ onset}$, defined in Figure S10). If the critical superconducting transition temperature ($T_c$) is defined as the temperature where the resistance approaches to zero, we find $T_c(0\ T) = 11.5$ K for $Fe_{1.13\pm0.06}Te$. These observations align with earlier works.[23, 43-44] The sharpness of the superconducting transition ($\Delta T_c = T_{c,\ onset} - T_c$) at zero field is ~ 1 K, as shown in the inset of Figure 4a, indicating the high crystalline quality of the grown sample.

Figure 4b depicts the temperature-dependent sheet resistance $R_s(B)$ of $Fe_{1.13\pm0.06}Te$ under different external perpendicular magnetic fields. As expected, as the field increases, the superconducting transition systematically shifts to lower temperatures. Note that, even in a magnetic field of 12 T, the $T_c$ (12 T) = 10.2 K is found to be very close to $T_c$ (0 T) = 11.5 K at zero magnetic field. This observation attests to the high quality of the sample. Besides observing the superconductivity in $Fe_{1.13\pm0.06}Te$ sample, we also detected superconductivity in $Fe_{1.43\pm0.07}Te$ sample, shown in Figure 4c and 4d. Interestingly, $Fe_{1.43\pm0.07}Te$ exhibits markedly distinct superconducting behaviors, as evidenced in Figure 4c. Firstly, it shows a quite broad superconducting transition $\Delta T_c$ = 2.6 K at $B = 0$ T, almost three times the value of $\Delta T_c = 1$ K in $Fe_{1.13\pm0.06}Te$, though both have a similar $T_{c,\,onset} \sim 12.6$ K. Secondly, a long tail is found to be superposed on the superconducting transition curves with the introduction of external magnetic fields, making the superconducting transition under magnetic fields very unusual. Finally, at the field of $B = 12$ T, the superconducting transition temperature is suppressed to 3.8 K, in stark contrast to 10.2 K in $Fe_{1.13\pm0.06}Te$, highlighting that Fe content can be used to tune superconductivity in $Fe_{1+y}Te$. It is well know that the sample thickness usually has an important effect on the superconductivity.[5, 23, 45-51] To examine the thickness effect on $T_c$ and $\Delta T_c$ of our CVD prepared $Fe_{1+y}Te$ samples, we fabricated several devices with different thicknesses. Figure S10 (Supporting Information) shows the thickness-dependent superconductivity for four typical devices (#1 ~ #4), where one can see that $T_{c,\,onset}$ shows a weak thickness dependence in our CVD-prepared $Fe_{1+y}Te$ superconductors. Together with the fact that the $T_{c,\,onset}$ observed here is consistent with prior studies on epitaxy films and single crystals[23, 43, 52], we can conclude that $T_{c,\,onset}$ of our CVD-samples is insensitive to the sample thickness. For $Fe_{1.13\pm0.06}Te$ samples, the thinnest device #1 exhibits a larger transition width $\Delta T_c$, which is ascribed to the enhanced thermal fluctuations [53-54] because its thickness approaches 2D limit. As for $Fe_{1.43\pm0.07}Te$, the broadening of superconducting transition (device #4) is interpreted as the vortex motion as discussed below.

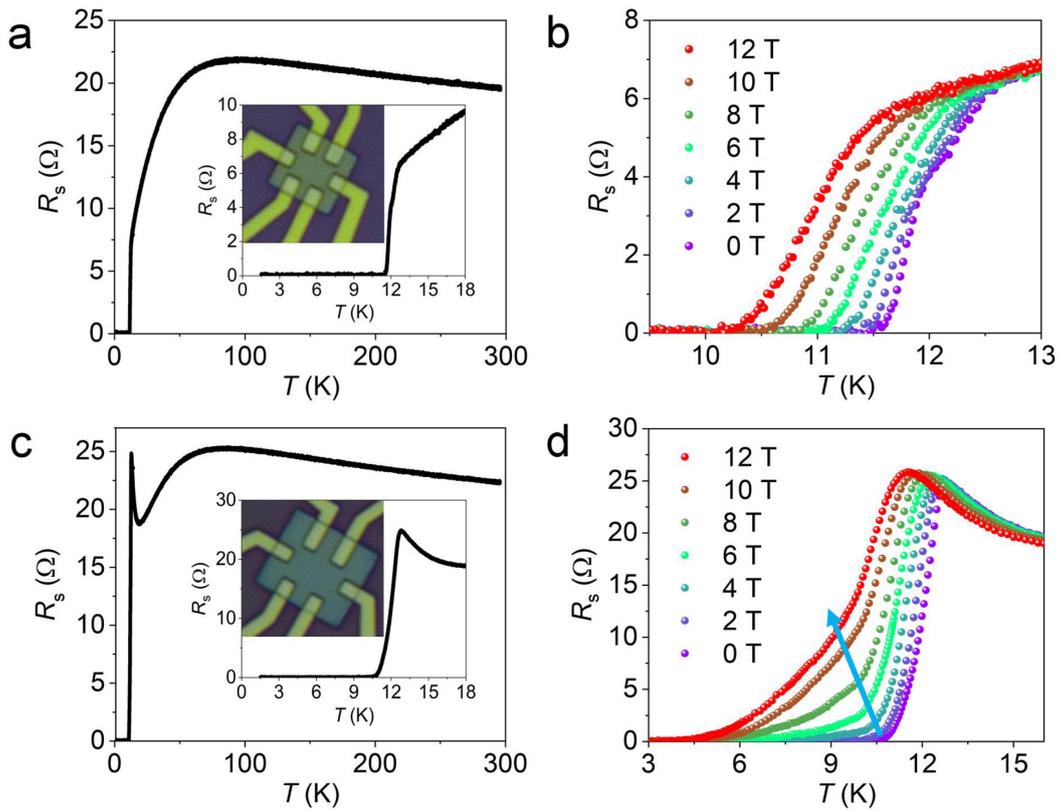

**Figure 4.** a, c) Temperature dependence of the sheet resistance ($R_s$) of tetragonal $Fe_{1.13\pm0.06}Te$ (Figure 4a) and $Fe_{1.43\pm0.07}Te$ (Figure 4c) samples. The inset shows an enlargement of $R_s$ within the temperature range of 18 K-1.5 K and the optical images of typical devices used in the low-temperature electrical transport measurements. b, d) Superconducting transition curves for $Fe_{1.13\pm0.06}Te$ (Figure 4b) and $Fe_{1.43\pm0.07}Te$ (Figure 4d) under varying external perpendicular magnetic fields. The blue arrow in (Figure 4d) marks the resistance kink of the $Fe_{1.43\pm0.07}Te$ sample.

### 2.4. The flux flow behavior of $Fe_{1.43\pm0.07}Te$.

If we take a closer look at the superconducting transition data $R_s(T)$ in Figure 4d, a resistance kink can be resolved in $R_s(T)$ curves denoted by the blue arrow, which is a signature of flux flow.[25, 55] It is known, for an ideal type-II superconductor, the vortices form the vortex lattice. With the introduction of impurity or disorder, the vortex lattice transforms into a glassy vortex state. As the temperature increases to a critical value characterized by $T_g$, the vortex glass melts into a vortex liquid state. Once the temperature exceeds $T_g$, the vortices start to move, which in turn results in the generation of resistance. From the definition of $T_g$ and $T_c$, one can easily find that they take the same value, though they have different physical meanings. $T_c$ depicts the transition from normal state to superconducting state, while $T_g$ depicts the vortex-glass to vortex-liquid transition. According to Fisher's theory,[56] the characteristics of vortices in non-ideal type-II superconductors can be discerned by examining the voltage-current ($V(I)$) behavior. When the temperature falls below $T_g$, the $V(I)$ curve displays a negative curvature at lower current values, indicating the presence of a vortex glass phase. Conversely, the $V(I)$ curve exhibits a positive curvature when $T > T_g$, which is a signature of the vortex liquid state. At $T = T_g$, the $V(I)$ curve exhibits a power law dependence with the following formula,

$$V \propto I^{(z+1)/2} \tag{1}$$

where $z$ is a material-dependent factor, and always takes a value of $4 < z < 6$.[57] To probe the vortex state in $Fe_{1.43\pm0.07}Te$ samples, $V(I)$ characteristic curves were taken at different temperatures, and **Figure 5**a plots the corresponding measurement results on a log-log scale. Indeed, a phase boundary denoted by the black dotted line between 10 K and 11 K can be discerned in Figure 5a, which separating $V(I)$ curves into two groups with opposite curvatures. Therefore, by fitting the black dotted line with equation (1), the parameter of $z$ is determined to be 3.86, consistent with theoretical expectation.[58] The above discussion demonstrates the occurrence of the vortex glass to vortex liquid phase transition.

Now, let's move to discuss the superconducting transition broadening in $Fe_{1.43\pm0.07}Te$ nanoflakes under magnetic fields. It is widely accepted that the thermally activated flux flow (TAFF) can lead to the broadening of the superconducting transitions, as reported in high-$T_c$ superconductor.[59] TAFF is usually associated with the pinning potential energy introduced by pinning centers in the superconductor. In $Fe_{1.43\pm0.07}Te$ sample with high Fe content, the excess Fe can act as the pinning centers and is responsible for the superconducting transition broadening, as demonstrated in Refs [25, 52, 60]. According to the TAFF theory,[61-62] the sheet resistance in the TAFF regime can be written as:

$$\ln R_s(T,H) = \ln R_0(H) - U_0(H)/T, \ln R_0(H) = \ln R_1 + U_0/T_c \qquad (2)$$

where $R_1$ is a constant and $U_0$ is the thermal activation energy. Consequently, it is possible to determine the activation energy $U_0$ of flux flow in the superconducting sample by equation (2). In Figure 5b, we present the plot of $\ln R_s$ vs. $1/T$ for the sample under different perpendicular magnetic fields. One can see clearly that there are two distinct parts where $\ln R_s$ shows linear dependences upon $1/T$ with different slopes, signifying the presence of two distinct thermal activation regimes, as denoted by '$h$' and '$l$' in Figure 5b. This is a typical behavior for dissipations induced by TAFF, and has been reported in MoGe thin films[63] and high-$T_c$ superconductor[64]. Extrapolating the fits to higher temperatures for the '$h$' regime, we find the fits cross at nearly the temperature $T_{c,\,onset} \sim 12$ K (see the crossing points $T_m$ in Figure 5b). From these two linear parts in each field, the activation energies $U_0^h$ and $U_0^l$ for the '$h$' and '$l$' temperature regimes can be extracted, which are plotted in Figure 5c on a semi-log scale and Figure 5d on a log-log scale, respectively. From the best fit, as denoted by the black solid lines, two different magnetic field dependent thermal activation behaviors are clearly revealed. At high temperatures, we find the thermal activation energy shows a logarithmic field dependence $U_0^h = 450 - 364 \ln H$, while $U_0^l$ follows a power-law behavior $U_0^l = 110.9 H^{-0.59}$ at low temperatures. These observations are consistent with the theoretical expectation for a phase transition from a 3D vortex liquid to 2D vortex liquid phase, and have been experimentally confirmed in copper-based superconductors.[64-65] Consequently, our electrical transport data demonstrate that, as the temperature increases, the sample undergoes a transition from a 3D vortex liquid state to a 2D vortex liquid state.

It is worth pointing out that the different superconducting behavior observed here are closely related to the Fe content. The parent compound FeTe has an antiferromagnetic order with an in-plane magnetic wave vector $(\pi, 0)$, in contrast, the superconducting $Fe_{1+y}Te_{1-x}Se_x$ sample exhibits an in-plane magnetic wave vector $(\pi, \pi)$. The theoretical calculation and experimental studies[66-68] have demonstrated that the excess Fe in

$Fe_{1+y}Te$ can provide local moments. The interaction between local moments and the plane Fe magnetism might either suppress the (π, 0) [69] order or stabilize the (π, π) order[70]. Meanwhile, the excess Fe serve as dopants and can provide charge carriers[68], which can also induce superconductivity as realized in the $Bi_2Te_3/Fe_{1+y}Te$[25, 71]. Therefore, excess Fe is necessary for the observation of superconductivity in $Fe_{1+y}Te$. However, as aforementioned, when much more Fe is introduced, they act as the pinning centers and broaden the superconducting transition. Hence, our results provide a promising knob to tune the superconducting properties of the 11 Fe-based superconductors.

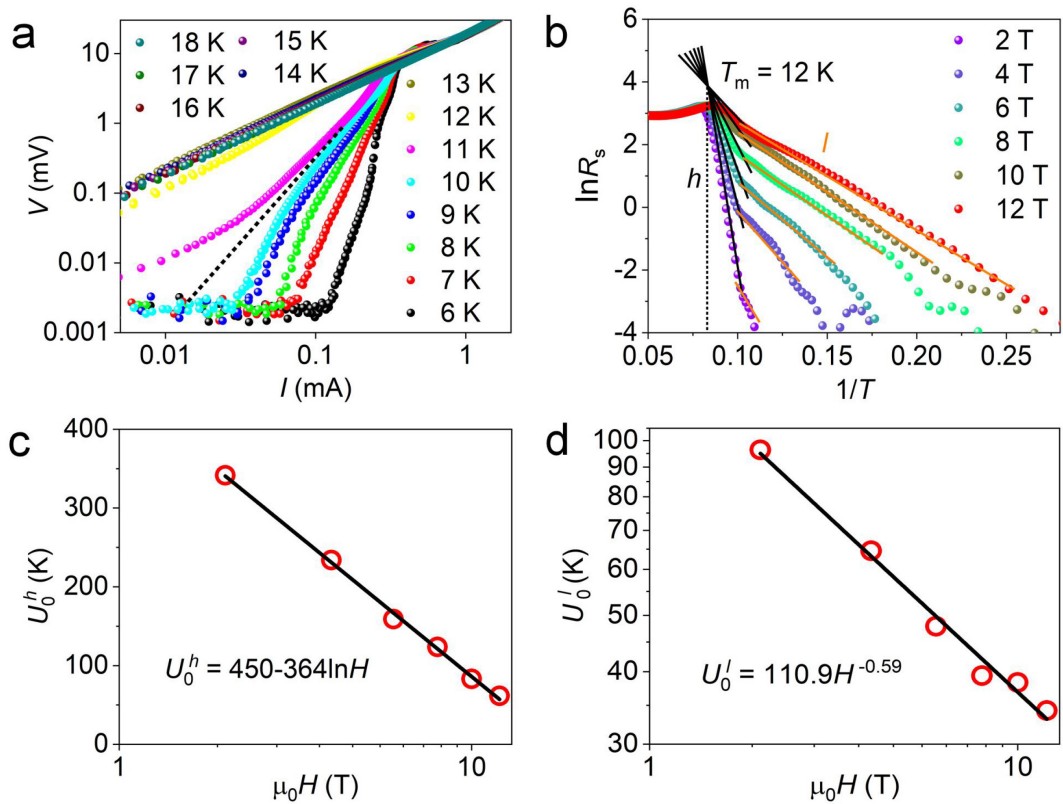

**Figure 5.** a) Voltage-current $V(I)$ characteristic curves using a log-log scale of $Fe_{1.43\pm0.07}Te$ nanoflakes at different temperatures. The dotted line represents the boundary where the curvature of the curve changes. b) ln $R_s$ vs. $1/T$ plots displaying two distinct linear regimes, from which the thermal activation energy $U_0$ can be extracted through data fitting. (c-d) Thermal activation energies $U_0^h$ and $U_0^l$ extracted from the slopes in the high temperature regime '$h$' (Figure 5c) and low-temperature

regime '*l*' (Figure 5d), respectively. The black solid lines represent the best fitting curves.

## 3. Conclusion

In summary, we have successfully synthesized the superconducting tetragonal $Fe_{1+y}Te$ nanoflakes via CVD method. The high quality of $Fe_{1+y}Te$ has been demonstrated by Raman, XPS and STEM, as well as low-temperature electrical transport measurements. Interestingly, a very sharp superconducting transition is observed in $Fe_{1.13\pm0.06}Te$, while the $Fe_{1.43\pm0.07}Te$ shows a relative broad superconducting transition. More importantly, a phase transition from a 3D vortex liquid state to a 2D vortex liquid state was observed in $Fe_{1.43\pm0.07}Te$. This work highlights that the CVD is an effective method to tune the superconductivity of Fe-based superconductors and the synthesized high-Fe content Fe-based superconductor provides an ideal platform for exploring the vortex dynamics.

## 4. Experimental Section

### CVD Synthesis of $Fe_{1+y}Te$ nanoflakes

The samples were grown in a 1-inch horizontal quartz tube furnace under atmosphere pressure. We used $FeCl_2$ partials (99.99%, Sigma Aldrich) and tellurium powder (99.99%, Aladdin) as precursors for growing $Fe_{1+y}Te$ nanoflakes. Approximately 0.1 g $FeCl_2$ precursor was placed in a porcelain boat, and a Si wafer with a 285 nm $SiO_2$ layer was placed on top of the porcelain boat with the polished side facing down. Another porcelain boat containing 5 g tellurium powder was placed in the upstream area at a temperature of about 400 °C. Before heating, ultra-high purity Ar was continuously blown into the quartz tube for 10 minutes to remove oxygen and moisture. During the whole process, a constant flow of $Ar/H_2$ was used as carrier gas through the furnace. The furnace was then heated to the growth temperature (520-600 °C) with a ramp rate of 50 °C/min and was held at this temperature for 3 minutes. Subsequently, the system was rapidly cooled down to room temperature.

### Sample Characterizations

The optical images were captured using an Olympus BX53 microscope. The Raman spectrum of the sample was acquired utilizing the WITEC Alpha 300 Raman system. The system was calibrated with the Raman peak at 520.7 cm$^{-1}$ of Si. The Raman spectrum and Raman mapping were carried out using a 532 nm laser as the incident light source. Elemental composition analysis was performed with XPS photoelectron spectroscope (Thermo Scientific K-Alpha).

## STEM experiment and characterizations

The silicon wafer containing the samples was first cut longitudinally using a FEI Helios G4 UC dual-beam microscope to obtain a cross-section of the samples. With regard to surface dust damage, we first polished the sample, then coated it with 20nm Pt and finally cut down a 100nm thick sample. Then FEI Themis Z equipped with a dual aberration corrector was used for in STEM and EDS data acquisition. This allowed us to obtain atomic structures with a spatial resolution of 60 pm at 300 KV and to map the EDS spectra in 10 minutes using a 4-detector setup. The data were then processed using Velox software.

## Device Fabrication and Electrical Transport Measurements

The Hall bar devices were fabricated on CVD-grown Fe$_{1+y}$Te nanoflakes using standard electron beam lithography. 5 nm Ti and 70 nm Au were used as contact electrodes. Electrical transport measurements of the prepared devices were conducted on the Oxford Instruments Teslatron TMPT system, which enables measurement conditions with a minimum temperature of 1.5 K and a maximum magnetic field strength of 14 T.

## Acknowledgements

The work was supported by the National Natural Science Foundation of China (grant No. 62174013, 92065203, 12104498, 61974120, 92265111 and 12104050), the funding Program of BIT (grant No. 3180012212214 and 3180023012204), the National Science Foundation for Distinguished Young Scholars (grant No. JQ23007), the Beijing Natural

Science Foundation (grant No. L233003), the National Basic Research Program of China from the MOST (grant No: 2022YFA1602803), the Strategic Priority Research Program of the Chinese Academy of Sciences (grant No. XDB33010300). This work was supported by the Synergic Extreme Condition User Facility and by the Innovation Program for Quantum Science and Technology (Grant No. 2021ZD0302601).

## Conflict of Interest

The authors declare no conflict of interest.

# Supporting Information

# High quality Fe$_{1+y}$Te synthesized by chemical vapor deposition with conspicuous vortex flow

*Lu Lv, Lihong Hu, Weikang Dong, Jingyi Duan, Ping Wang, Peiling Li, Fanming Qu, Li Lu, Zimeng Ye, Junhao Zhao, Jiafang Li, Fang Deng, Guangtong Liu\*, Jiadong Zhou\*, Yanfeng Gao\**

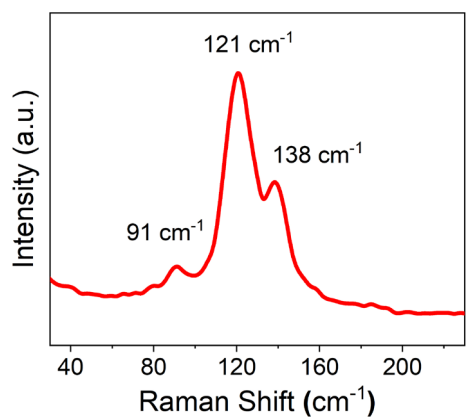

**Figure S1.** Raman spectrum of a FeTe nanoflake under the 532 nm excitation. The sample shows the typical Raman peaks of tetragonal phase, featured by the 91 cm$^{-1}$, 121 cm$^{-1}$ and 138 cm$^{-1}$ peaks.

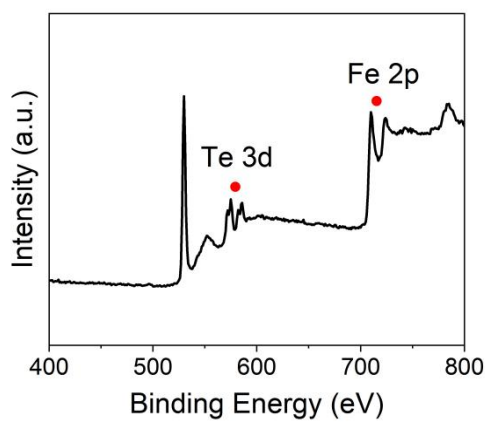

**Figure S2.** Full XPS spectrum of the obtained FeTe nanoflakes. The observed peak

positions in the spectrum align with the binding energies of the Fe and Te elements, confirming the existence of the constituent components in the FeTe samples.

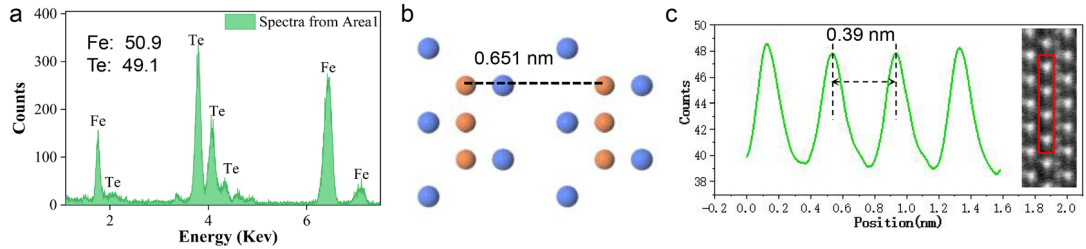

**Figure S3.** a) Quantified elemental analysis of a FeTe nanoflake. b) Side view of FeTe crystal structure. Brown and blue colors represent Fe and Te atoms, respectively. c) Distribution diagram of atomic spacing along the a-axis between two Te atoms in frame of Figure 2c.

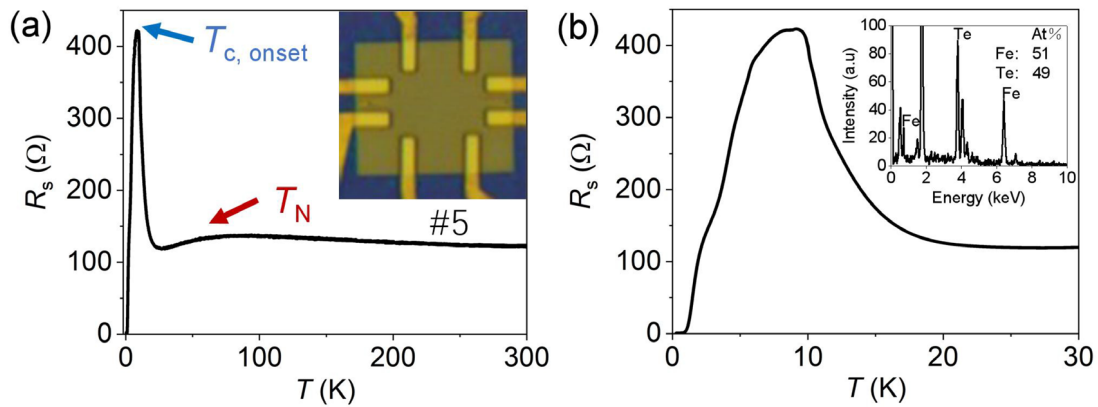

**Figure S4.** a) Temperature dependence of sheet resistance of FeTe sample. b) The expanded temperature region close to the superconducting transitions. The inlet shows the EDX data of sample #5.

Figure S4 shows the temperature dependence of sheet resistance $R_s(T)$ for FeTe (#5). With decreasing temperature, $R_s$ increases slightly and forms a broad peak around $T_N$ = 75 K, which is regarded as the Neel temperature where the system transitions from the paramagnetic (PM) state to antiferromagnetic (AFM) state[1,2]. Then, $R_s$ increases steeply followed by a superconducting transition at ~ 9 K (Figure S4a). Furthermore, the sample ultimately transitions into the zero-resistance state below 1 K. The superconducting behavior in FeTe has been reported in reference [3] which shows a similar $T_{c,\ onset}$ with our sample. From the inset of Figure S4b, we can see that the atomic ratio in of Fe:Te is close to 1:1.

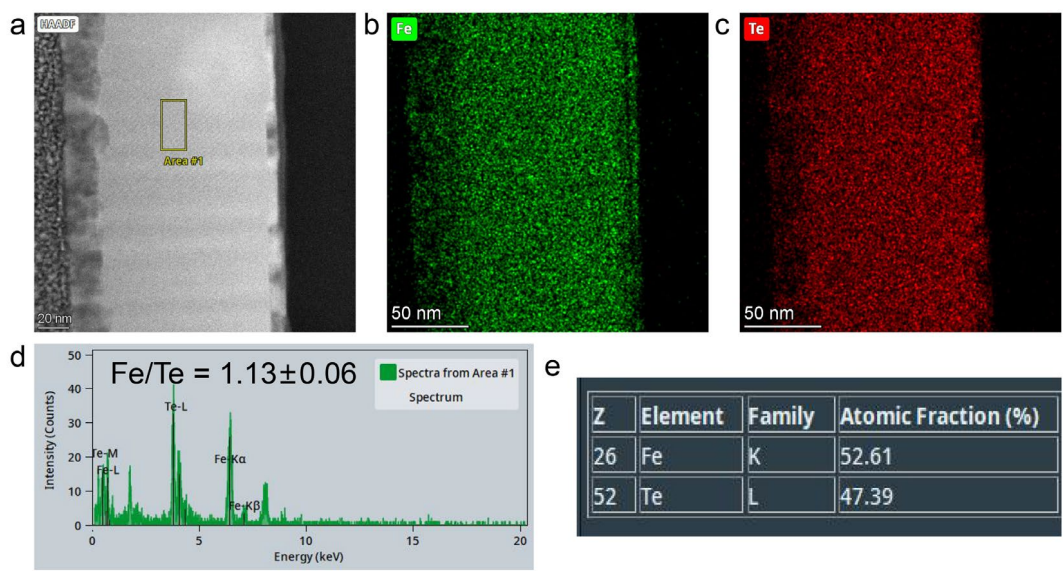

**Figure S5.** a) Low-magnified STEM image of a $Fe_{1+y}Te$ nanosheet. b-c) EDX mapping of Fe and Te elements in $Fe_{1+y}Te$ nanosheet, respectively. The low contrast on the surface layer of the nanosheet (i.e. the left and right layers of the slabs) is attributed to the damaged sample caused by the force of FIB's Ga-ion. d) Quantified EDX elemental analysis of a Fe1+yTe nanoflake. The margin of relative error in EDX analysis is typically within 5%[1]. Consequently, the actual composition of the $Fe_{1.13}Te$ sample may fall within the range of $Fe_{1.13\pm0.06}Te$.

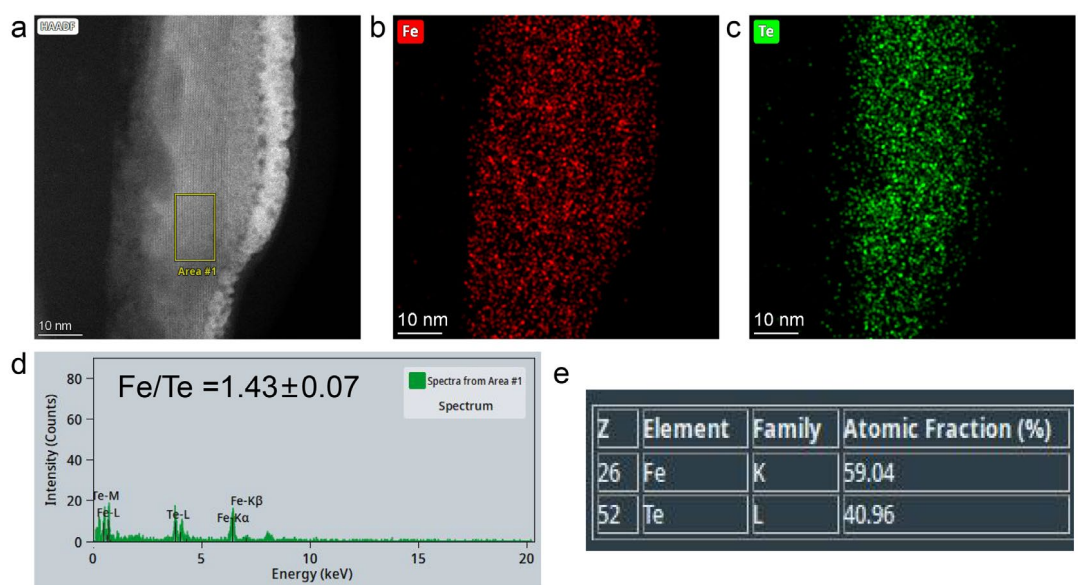

**Figure S6.** a) Low-magnified STEM image of a Fe$_{1+y}$Te nanosheet. b-c) EDX mapping of Fe and Te elements in Fe$_{1+y}$Te nanosheet, respectively. The low contrast on the surface layer of the nanosheet (i.e. the left and right layers of the slabs) is attributed to the damaged sample caused by the force of FIB's Ga-ion. d) Quantified EDX elemental analysis of a Fe$_{1+y}$Te nanoflake, the Fe/Te ratio is measured to be 1.43. Due to the margin of relative error in EDX analysis within 5%[1], the actual composition of the Fe$_{1.43}$Te sample may fall within the range of Fe$_{1.43\pm0.07}$Te.

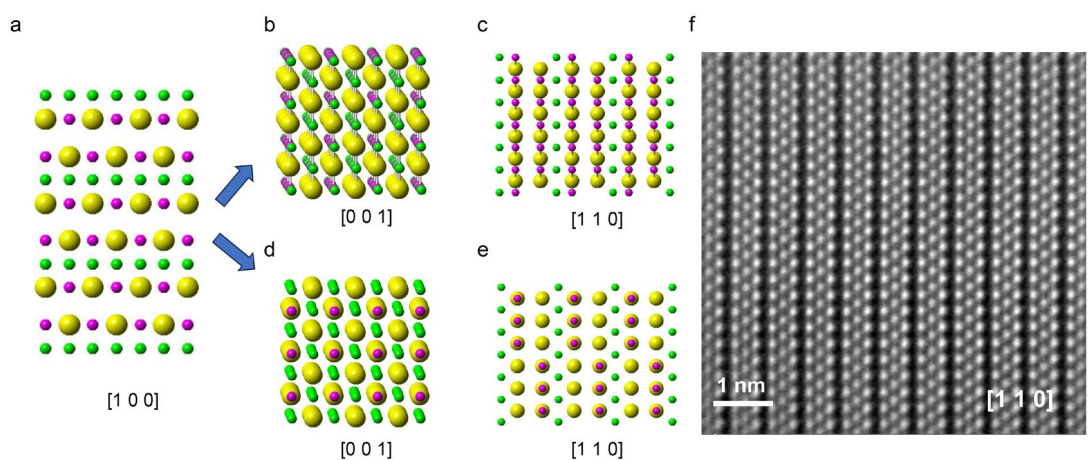

**Figure S7.** a) Atom model of Fe$_{1+y}$Te crystal along the [1 0 0] crystal direction. b, d) Atom model of Fe$_{1+y}$Te crystal along the [0 0 1] crystal direction. c, e) Atom model of Fe$_{1+y}$Te crystal along the [1 1 0] crystal direction. f) STEM image of a Fe$_{1+y}$Te sample

along the [1 1 0] crystal direction.

If the excess Fe atom shown as pink color is located between Te-Te atoms in the [1 0 0] direction in Figure R1a, the relative [1 1 0] and [0 0 1] direction would seen two different atomic structure as shown in Figure S7b-c and Figure S7d-e respectively, where the excess Fe atoms locate in Fe column or Te column shown in Figure S7b or Figure S7d in [0 0 1] which difficult to confirm due to its low contrast. However, when the crystal are tilt to [1 1 0] direction, we could observe huge different atomic structure as shown in Figure S7c or Figure S7e, we acquire the relative HAADF-STEM image in [1 1 0] shown in Figure S7f that the excess Fe couldn't be observe confirming the actual $Fe_{1+y}Te$ are corresponding to Figure S7d and Figure S7e.

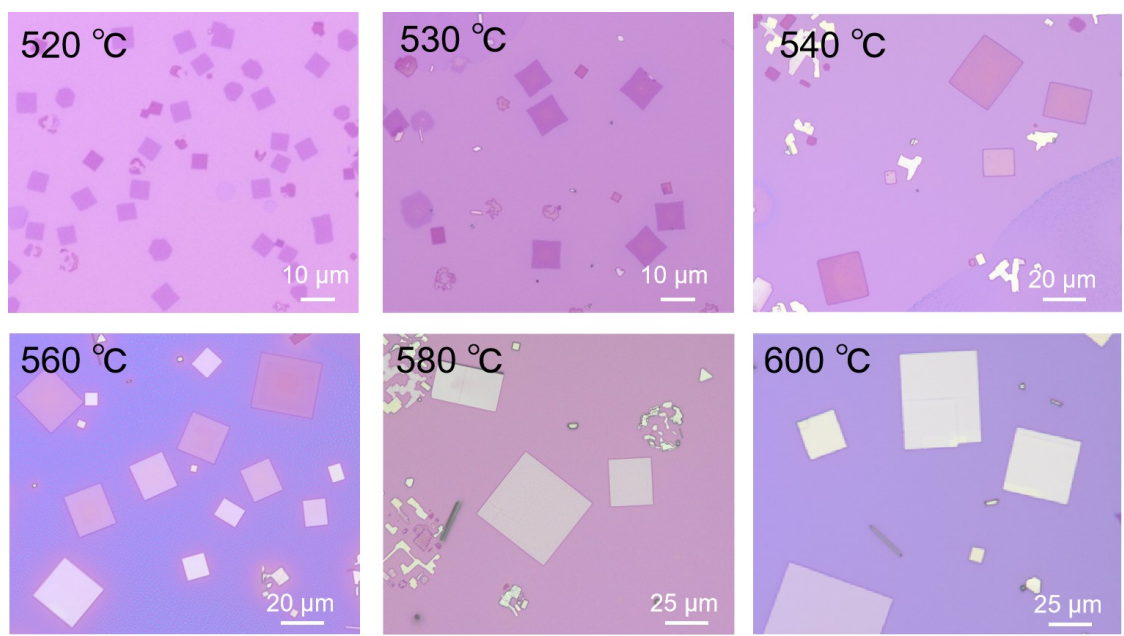

**Figure S8.** The optical images of FeTe nanoflakes grown on the $SiO_2$/Si substrate at different growth temperatures. The mass ratio of $FeCl_2$ and Te precursors used in the experiments is ~2:1. The thickness and the size increase with increasing temperature.

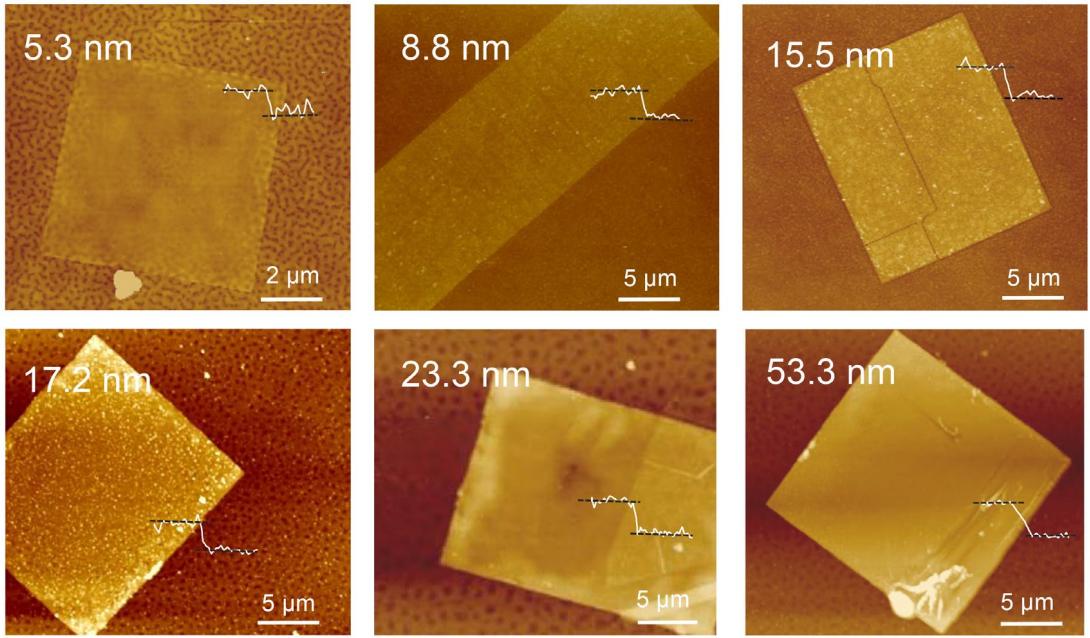

**Figure S9.** AFM images of tetragonal FeTe nanoflakes with different thicknesses (5.3 nm, 8.8 nm, 15.5 nm, 17.2 nm, 53.3 nm, respectively).

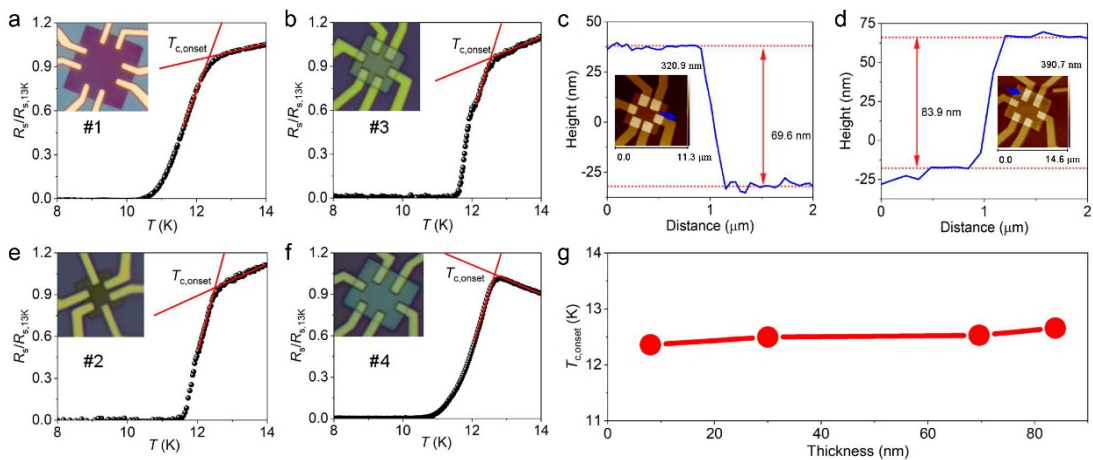

**Figure S10.** Thickness-dependent superconductivity in $Fe_{1.13\pm0.06}Te$ (a)-(c) and $Fe_{1.43\pm0.07}Te$ (d) flakes ranging from 8 nm to 84 nm. $T_{c,\,onset}$ is defined as the onset temperature of the superconducting transition, obtained by the intersection of the extrapolation of normal-state resistivity and superconducting transition. e, f) AFM height profiles of devices #3 and #4 with the thickness of 69.6 nm and 83.9 nm. Inset: AFM images of devices #3 and #4. g) Thickness dependent $T_{c,\,onset}$.